\documentclass[reprint,aps,pre,groupedaddress,showkeys]{revtex4-1}
\usepackage[T1]{fontenc}
\usepackage[utf8]{inputenc}
\usepackage[english]{babel}
\usepackage{amsmath}
\usepackage{amsfonts}
\usepackage{amssymb}
\usepackage[colorlinks=true,linkcolor=blue,urlcolor=blue,citecolor=red]{hyperref}
\usepackage{lineno}
\usepackage{braket}
\usepackage{wrapfig}
\usepackage{todonotes}
\usepackage{graphicx}
\usepackage{float}
\usepackage{natbib}
\usepackage[left=2cm,right=2cm,top=2cm,bottom=2cm]{geometry}

\newcommand{\mat}[1]{\mathbb{#1}}
\newcommand{\mi}{\mathrm{i}}

\begin{document}
    \title{Dynamical analysis of mass-spring models using Lie algebraic methods}
    \author{Alejandro R. Urz\'ua}
    \email[Corresponding author: ]{arurz@inaoep.mx}
    \affiliation{Instituto Nacional de Astrof\'{\i}sica, \'Optica y Electr\'onica, Calle Luis Enrique Erro No. 1, Santa Mar\'{\i}a Tonantzintla, Puebla, 72840, Mexico}
    \author{Ir\'an Ramos-Prieto}
    \email[Corresponding author: ]{iranrp123@gmail.com}
    \affiliation{Instituto Nacional de Astrof\'{\i}sica, \'Optica y Electr\'onica, Calle Luis Enrique Erro No. 1, Santa Mar\'{\i}a Tonantzintla, Puebla, 72840, Mexico}
    \affiliation{Instituto de Ciencias F\'{\i}sicas, Universidad Nacional Aut\'onoma de M\'exico, Apartado Postal 48-3, 62251 Cuernavaca, Morelos, M\'exico}
    \author{Francisco Soto-Eguibar}
    \affiliation{Instituto Nacional de Astrof\'{\i}sica, \'Optica y Electr\'onica, Calle Luis Enrique Erro No. 1, Santa Mar\'{\i}a Tonantzintla, Puebla, 72840, Mexico}
    \author{H\'ector Moya-Cessa}
    \affiliation{Instituto Nacional de Astrof\'{\i}sica, \'Optica y Electr\'onica, Calle Luis Enrique Erro No. 1, Santa Mar\'{\i}a Tonantzintla, Puebla, 72840, Mexico}

    \date{\today}

\begin{abstract}
    The dynamical analysis of vibrational systems of masses interconnected by restitution elements each with a single degree of freedom, and different configurations between masses and spring constants, is presented. Finite circular and linear arrays are studied using classical arguments, and their proper solution is given using methods often found in quantum optical systems. We further study some more complicated arrays where the solutions are given by using Lie algebras.
\end{abstract}

    %\pacs{Valid PACS appear here}
    \maketitle

\section{Introduction}\label{intro}
The dynamical description of systems with coupled subelements is a
well known subject in classical mechanics literature
\cite{Meirovitch1986,landauM,Kotkin1980}. It is a typical task to
take the stated system of dynamical coupled equations, that emerge
from the equilibrium analysis, and solve them with some
differential equations or eigenvalue techniques. A simple model
that describes a chain of classical particles (atoms) harmonically
coupled with their nearest neighbors and subjected to a periodic
on-site (substrate) potential has become in recent years one of
the fundamental and universal models of low dimensional nonlinear
physics. In spite of the fact that a link with the classical model
is not often stated explicitly in many applications, many kind of
nonlinear problems involving the dynamics of discrete nonlinear
chains are in fact based on the classical formulation introduced
in the papers by Ya. Frenkel and T. Kontorova
\cite{YaFrenkel1938,T.A.Kontorova1938,T.A.Kontorova1938a,Braun1998,Allen1998},
who suggested to use this kind of nonlinear chain to describe,
in the simplest way, the structural dynamics of a crystal lattice
in the vicinity of the dislocation core
\cite{Braun1998,Allen1998}.\\
Furthermore, the study of these particular systems, that
are intrinsically classical due to their macroscopic mechanical
nature, can be linked to the study of a propagated light field
under a waveguide array, which in turn can also be described by
classical means. Some work has been made on photonic lattices to
study the analogy between quantum systems and classical light
propagation \cite{Makris2006,Keil2011,PerezLeija2010}. Under this
scheme, the system of equations that describes the time evolution
of the oscillation amplitude can be written in terms of operators
that fulfill certain known commutation relations from quantum
mechanics. In this sense, the system of equations of a tight-
binding model to first neighbors and periodic boundary conditions
can be written using the discrete Fourier transform
\cite{PerezLeija2016}. On the other hand, linear finite arrays in
both paradigms, light or mechanical, exhibit a similitude when the
same methods to obtain solutions and insights are applied. The
emulation of quantum mechanical properties with classical
propagated light is well known \cite{Makris2006,PrezLeija2012},
following here that the nearest neighbor interaction of mechanical
systems mimics some of the features encountered in the classical
light counterpart.\\
The goal of this manuscript is to take a few well-known classical
physical systems and fully solve them with techniques found often
in the description and solution of quantum mechanical problems. In
Section \ref{circarray}, we start with the analysis of a finite
set of masses connected to each other by springs with constants of
harmonic restitution following Hook's law, constrained to move
with a single degree of freedom in a circle. The set of differential
equations that describe the system has a periodic boundary
condition, which allows the first and last element to be coupled.
In Section \ref{linear_array}, we restate the problem removing the
boundary condition of the circular array and keeping only a finite
chain fixed at both or one of the edges. In this class of arrays
we distinguish when all the restitution coefficients and masses
are equal \ref{lineq}, which leads to an analytical solution given
by Chebyshev polynomials of the second kind. The rise of traveling
normal waves is observed. If the restitution coefficients follows
some other law in function of the position of the masses, then
other phenomena are present. We engineer an interaction matrix
\ref{linkr} where his diagonalization is given in terms of
Kravchuk functions that, in turn, are solutions of the discrete
and finite harmonic oscillator of $\mathfrak{su}(2)$
\cite{Atakishiyev2008,ata}. Here, the solution is also oscillatory
but with persistence of the poles and nodes presented by the
Kravchuk functions that are in the core hypergeometrical
functions. We observe bouncing of amplitudes that become rapidly a
complete interference patterns when the propagation time is
sufficiently large, giving no recombination nor recovering of
initial conditions. We conclude with the approach and solution of
a problem that relates to the masses and springs through binomial
coefficients \ref{linfac}. Here, although the analytical solution
exists and belongs too to the realm of the well-known algebra
$\mathfrak{su}(2)$, but it is somewhat not easy to calculate explicitly,
and instead numerical results are presented.

\section{Circular finite array}\label{circarray}
\begin{figure}[H]
    \begin{center}
        \includegraphics[width=.65\linewidth]{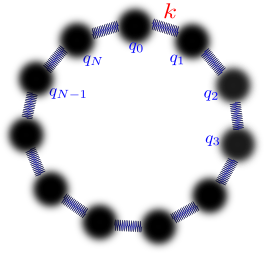}
    \end{center}
    \caption{Scheme of the finite circular array of masses in clockwise order. Because of the arrangement, the system has circular symmetry and it maps onto itself when a rotation by an arbitrary angle is made. Furthermore, the label of the masses obeys a modular count when $q_{N+s}=q_{s-1}$ for $s=1,2,\cdots$.}
    \label{fig01}
\end{figure}

\begin{figure}[H]
    \begin{center}
        \includegraphics[width=\linewidth]{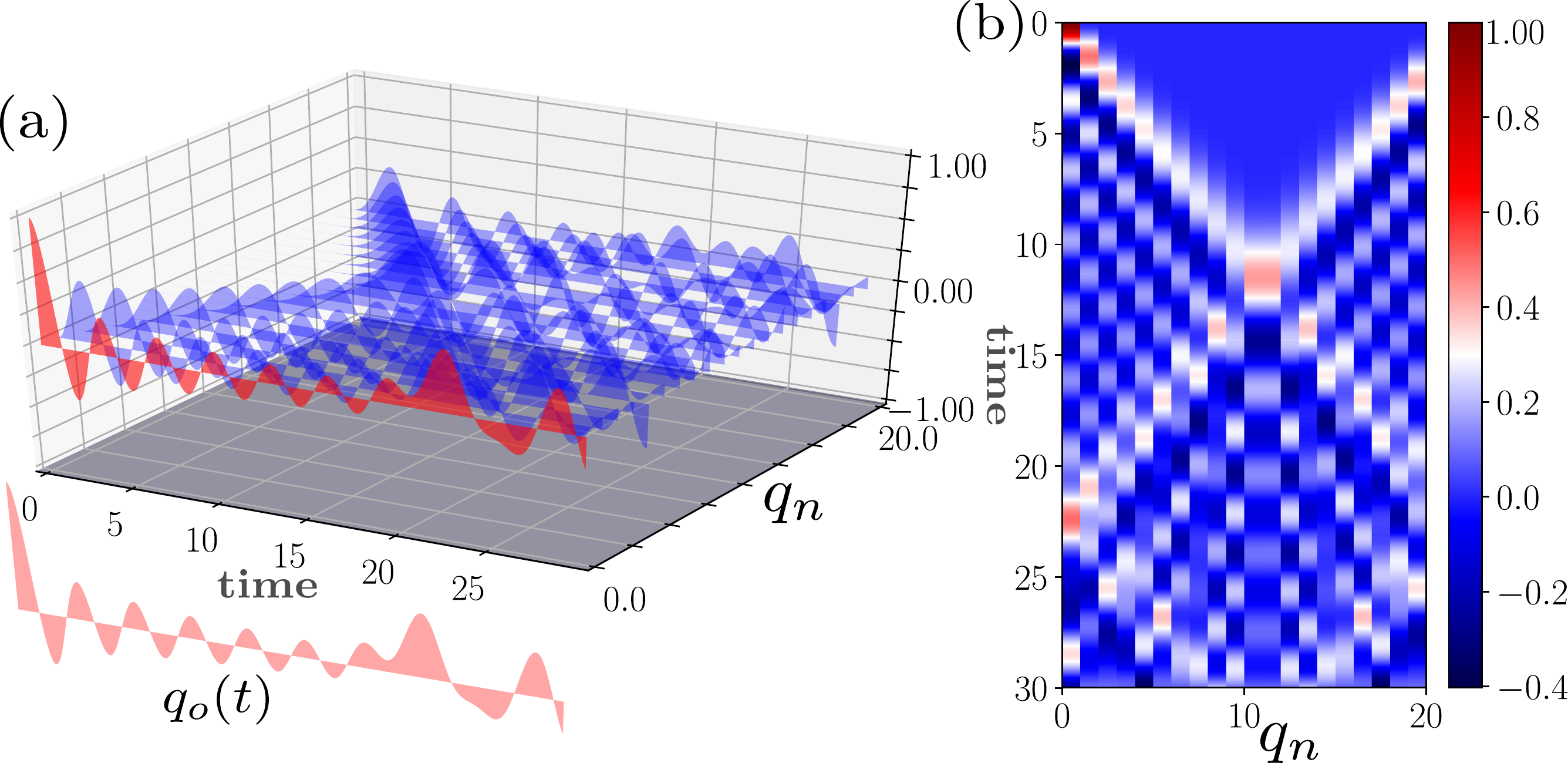}
    \end{center}
    \caption{Temporal evolution of the position of each mass in the circular array. The initial condition has unitary amplitude at $\mathbf{Q}(0)=\ket{0}$. We set $N=30$. It is worth to notice the similar behavior with the one in waveguide systems of classic light.}
    \label{fig02}
\end{figure}

Let us consider the interaction between a finite set of masses labeled by $m_{j}, \; j=0,1,2,...,N$, where the dimension of the set is $d_{N}=N+1$. Geometrically arranged in a circle, the interaction of $m_{j}$'s is mediated by springs with equal restitution constants $k$, as it is shown schematically in Fig. \ref{fig01}. The position of every single mass is labeled by the canonical coordinate $q_{j}(t), \; j=0,1,2,...,N$. Taking $m_{j}=1$ for every $j$ in the set, the coupled equations of motion of this system is given by \cite{Meirovitch1986}
\begin{align}\label{eq01}
\ddot{q}_{0}+k(2q_{0}-q_{N}-q_{1})&=0, \nonumber\\
\ddot{q}_j+k(2q_j-q_{j+1}-q_{j-1})&=0,\\
\ddot{q}_N+k(2q_N-q_{0}-q_{N-1})&=0, \nonumber
\end{align}
where for the sake of simplicity, we drop the time dependence of the space coordinates, $q_{j}(t)\equiv q_{j}$.
\noindent It is straightforward to cast the set of differential equations \eqref{eq01} onto the matrix form
\begin{equation}\label{eq02}
\ddot{\mathbf{Q}}=\mathbb{M} \mathbf{Q},
\end{equation}
where $\mathbb{M}$ is the tridiagonal real matrix plus bounded corners, with dimensions $d_{N}^{2}$, explicitly given by
\begin{equation}\label{eq03}
\mathbb{M}_\mathrm{c}=k
\begin{pmatrix}
-2 & 1 & 0 & 0 & \cdots & 1\\
1 & -2 & 1 & 0 & \cdots & 0\\
0 & 1 & -2 & 1 & \cdots & 0\\
0 & 0 & 1 & -2 & \ddots & \vdots\\
\vdots & \vdots & \vdots & \ddots & \ddots & 1\\
1 & 0 & 0 & \cdots & 1 & -2
\end{pmatrix},
\end{equation}
being $\mathbf{Q}$ the $d_{N}$ column vector of canonical coordinates $q_{j}$'s
\begin{equation}\label{eq04}
\mathbf{Q}=
\begin{pmatrix}
q_{0} & q_{1} & q_{2} & \cdots & q_{N}
\end{pmatrix}^{T},
\end{equation}
where clearly $q_{j}$ is time dependent, so $\mathbf{Q}\equiv\mathbf{Q}(t)$.\\
Because we are dealing with a set of second order differential equations in \eqref{eq01}, we need to establish initial conditions, one set in the positions $\mathbf{Q}(0)$, and other set in the velocities $\dot{\mathbf{Q}}(0)$. A careful looking onto the matrix form \eqref{eq03} suggest an \emph{ansatz} for the initial value problem \eqref{eq02}; mainly, it needs to be a continuous differentiable real function with no parity associated, thus we propose it to be
\begin{equation}\label{eq05}
\mathbf{Q}\left( t  \right) =\cosh(t\sqrt{\mathbb{M}_\mathrm{c}})\mathbf{Q}(0),
\end{equation}
which can be verified as a legal solution by direct substitution in Eq. \eqref{eq02}.\\
Now is time to restructure the problem \eqref{eq02} in order to use some of the methods encountered in quantum optics. First, we notice that the coefficient matrix \eqref{eq03} has a tridiagonal form and the corners occupied with ones, which suggests to use the well known London operators $\{\mat{V},\mat{V}^{\dagger}\}$, whose matricial representation is given by \cite{London1926,PerezLeija2016}
\begin{equation}
\mat{V}=
\begin{pmatrix}
0 & 1 & 0 & 0 & \cdots & 0 \\
0 & 0 & 1 & 0 & \cdots & 0\\
0 & 0 & 0 & 1 & \cdots & 0\\
0 & 0 & 0 & 0 & \ddots & \vdots \\
\vdots & \vdots & \vdots & \ddots & \ddots & 1 \\
1 & 0 & 0 & \cdots & 0 & 0
\end{pmatrix}.\label{eq06}
\end{equation}
It can be shown that the matrix $\mat{V}$ obeys the spectral decomposition $\mat{V}={\mat{F}}\Lambda {\mat{F}}^{\dagger}$ \cite{moya2011differential}, where $\mat{F}$ is the discrete Fourier transform given by the Vandermonde confluent matrix \cite{PerezLeija2016,MoyaCessa2018}
\begin{equation} \label{eq07}
\mathbb{F}= \frac{1}{\sqrt{N+1}}
\begin{pmatrix}
1 & 1 & 1 & \cdots & 1 \\
\lambda_{0} & \lambda_{1} & \lambda_{2} & \cdots & \lambda_{N}\\
\lambda_{0}^{2} & \lambda_{1}^{2} & \lambda_{2}^{2} & \cdots & \lambda_{N}^{2}\\
\lambda_{0}^{3} & \lambda_{1}^{3} & \lambda_{2}^{3} & \cdots & \lambda_{N}^{3} \\
\vdots & \vdots & \vdots & \ddots & \vdots\\
\lambda_{0}^{N} & \lambda_{1}^{N} & \lambda_{2}^{N} & \cdots & \lambda_{N}^{N}
\end{pmatrix}
\end{equation}
with
\begin{equation} \label{eq08}
\lambda_{j} = \exp\left(\mi\frac{2\pi}{N+1}j\right),  \qquad j=0,1,2,...,N
\end{equation}
the $j$th root of the unity, and
\begin{equation}
\Lambda=
\begin{pmatrix}
\lambda_{0} & 0 & 0 & \cdots & 0\\
0 & \lambda_{1} & 0 & \cdots & 0\\
0 & 0 & \lambda_{2} & \cdots & 0\\
\vdots & \vdots & \vdots & \ddots & \vdots\\
0 & 0 & 0 & \cdots & \lambda_{N}
\end{pmatrix}
\label{eq09}
\end{equation}
as the eigenvalue diagonal matrix.\\
Using the former equations \eqref{eq06}, \eqref{eq07} and \eqref{eq08}, we can substitute in \eqref{eq05} and simplify the proposed solution as
\begin{align}\label{eq10}
\nonumber
\mat{Q}\left(t\right) &=\cosh\left[t\sqrt{k(\mat{V}+\mat{V}^{\dagger})-2k\mat{I}}\right] \mathbf{Q}(0)
\\  &
=\mat{F}\cosh\left[t\sqrt{k(\Lambda+\Lambda^*)-2k\mat{I}}\right]\mat{F}^{\dagger}\mathbf{Q}(0),
\end{align}
where it is worth to notice that the terms inside the square root of \eqref{eq10} are a sum of pure diagonal matrices.\\
Taking the basis vectors to be a Kronecker basis of dimension
$d_{N}$ of the form $\ket{j}:=\delta_{n,j}$ for $n=0,1,2,\dots,N$;
that is, defined $j$, the ket $\ket{j}$ is a vector that has zeros
everywhere, except when $n=j$; when the Hilbert space dimension is
infinite, these states are known as a Fock states in the quantum
mechanical realm. These kets can be used to define the Fourier
kernel \eqref{eq07} as an operator $\mat{F}$ in terms of the outer
product of their elements as \cite{PerezLeija2016},
\begin{equation}\label{eq11}
\mat{F} := \frac{1}{\sqrt{N+1}}\sum_{n,m=0}^{N}\exp\left[\mi\frac{2\pi mn}{N+1}\right] \ket{m}\bra{n},
\end{equation}
for which can be easily verified the inverse property $\mathbb{F}^{\dagger}=\mathbb{F}^{-1}$. Also the hyperbolic cosine in \eqref{eq09} can be expressed, with the help of Eqs. \eqref{eq08} and \eqref{eq07}, as
\begin{align}\label{eq12}
\nonumber
\cosh&\left[  t\sqrt{k(\Lambda+\Lambda^*)-2k\mathbb{I}}\right]=
\\ &
=\sum_{m=0}^{N}\cos\left[ 2 t \sqrt{k}\sin\left(\frac{\pi m}{N+1}\right)\right] \ket{m}\bra{m}.
\end{align}
With this restructured problem, we need only to establish the
initial conditions on the positions and leave the initial
velocities unknown. The initial conditions read $\mathbf{Q}(0)=w\ket{l}$ in terms of the orthonormal basis $\ket{l},\; l\in\{0,1,2,\cdots,N\}$, and it is a vector in a finite Hilbert space of dimension $N+1$. This means that the $l$-th mass in
the circle is \emph{excited} with an excitation weight $w$; in
other words, an amplitude of initial perturbation. We are now in
place to forward the proposed solution \eqref{eq05} onto a closed
analytical form, with the help of \eqref{eq10}, \eqref{eq11},
\eqref{eq12} and the initial condition. Remembering that
$\braket{r|s}=\delta_{r,s}$, we arrive to
\begin{align}\label{eq13}
\mathbf{Q}(t)=&\frac{w}{N+1}\sum_{n,m=0}^{N}\cos\left[\frac{2\pi m (n-l)}{N+1}\right] \nonumber
\\ &
\times\cos \left[ 2 t \sqrt{k}\sin\left(\frac{\pi m}{N+1}\right) \right] \ket{n},
\end{align}
which means that for every single solution in the vector $\mathbf{Q}(t)$, we obtain the solely dynamic dictated for every mass in the array by
\begin{align}\label{eq14}
q_{n}(t)=&\frac{w}{N+1}\sum_{m=0}^{N}\cos\left[\frac{2\pi m (n-l)}{N+1}\right]
\nonumber \\ &
\times\cos \left[ 2 t \sqrt{k}\sin\left(\frac{\pi m}{N+1}\right) \right].
\end{align}
In Figure \ref{fig02}, we plot the time evolution \eqref{eq14} for $N=30$ and an initial condition $\mathbf{Q}(0)=\ket{1}$. As can be seen, the general mechanical behavior is oscillatory. It is important to notice, that because of the periodic boundaries in the array, part of the initial amplitude is transmitted from the $q_{j}$ mass to the neighbor $q_{j+N}$; in this case some of the initial amplitude in $q_{0}$ at time $t_{0}$ is transmitted to $q_{30}$ at a time $t+\Delta t$, explaining that the coupled evolution plot have interference at some time $t_{i}$. It is equivalent to say, because the circular arrange of the masses, that the initial condition on $\ket{0}$ when the system evolves, looks like two initial conditions, one at $t_{0}$ and the other, retarded, at time $t+\Delta t$.

\section{Linear finite array of $N+1$ masses} \label{linear_array}
\begin{figure}[H]
    \begin{center}
        \includegraphics[width=.95\linewidth]{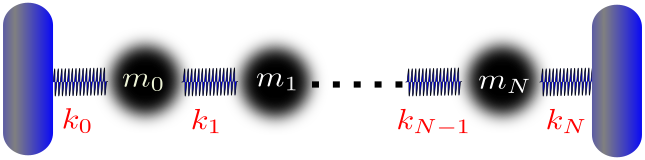}
    \end{center}
    \caption{Scheme of the linear finite array. There are $N$ masses and $N+1$ restitution elements, giving $2N+1$ entities in the array. The boundary conditions are not strictly of physical or material type, just mathematical constraints to keep the problem well defined.}
    \label{figmk}
\end{figure}

We now consider the task to determine the dynamic evolution of a linear finite array of masses $m_j, \; j=0,1,2,...,N$, with restitution elements $k_j, \; j=0,1,2,3,...,N$, as seen above in Section \ref{circarray}, and which schematic representation is shown in Fig. \ref{figmk}. In this case, the system of ordinary differential equations that governs the evolution is given by \cite{Meirovitch1986}
\begin{align}\label{eq15}
        m_0 \ddot{q}_0+\left(k_0+k_1 \right)q_0-k_1 q_1&=0,\nonumber \\
        m_j \ddot{q}_j+\left( k_j+k_{j+1}\right) q_j
        -k_{j+1} q_{j+1}-k_j q_{j-1}&=0,\\
        m_N \ddot{q}_N+\left(k_N+k_{N+1} \right) q_N -k_N q_{N-1}&=0.\nonumber
\end{align}
This system can be packed into the matrix representation
\begin{equation}\label{eq16}
\ddot{\mathbf{Q}}(t)=\mathbb{M}_\mathrm{L}\mathbf{Q}(t),
\end{equation}
where $\mathbb{M}_{L}$ is represented in the finite basis $\ket{l}$ as
\begin{align}\label{eq17}
\mathbb{M}_\mathrm{L}&=-\sum_{j=0}^{N}\frac{k_j+k_{j+1}}{m_j}\ket{j}\bra{j}
\nonumber \\ &
+\sum_{j=0}^{N-1}\frac{k_{j+1}}{m_{j+1}}\ket{j+1}\bra{j}
+\sum_{j=0}^{N-1}\frac{k_{j+1}}{m_{j}}\ket{j}\bra{j+1}
\end{align}
and
\begin{equation}\label{eq18}
\mathbf{Q}(t)=\sum_{j=0}^{N}q_j\left(t \right) \ket{j}.
\end{equation}
Using the ansatz previously given, the proper solution of \eqref{eq16} can be stated as
\begin{equation}\label{eq19}
\mathbf{Q}(t)=\cosh\left(t \sqrt{\mathbb{M}_\mathrm{L}} \right)\mathbf{Q}\left( 0 \right).
\end{equation}
It is clear that the real difficulty lies on two facts of the matrix function $\cosh\left(t \sqrt{\mathbb{M}_\mathrm{L}} \right)$;  first, we need to evaluate the matrix $\mathbb{M}_{L}$ at time $t$; second, this evaluation need to operate onto the initial condition on the right. The last issue can be simple tackle in the case when $\mathbb{M}_{L}$ can be diagonalized, this lets us obtain a simple form of the solution. The task now is to examine some cases where the matrix could be diagonalized in function of the nature of $m_{j}$ and $k_{j}$.

\subsection{$m_{j}$ and $k_{j}$ equal to one for all $j$'s}\label{lineq}
It is straightforward to take all masses and restitution elements as one; that is $m_{j} = 1$ and $k_{j} = 1$ for all $j\in{0,1,2,\cdots N}$ as they exhibit no dependence across the array positions nor the time parameter. Rewriting \eqref{eq17} with the new conditions, we obtain
\begin{equation}\label{eq20}
\mathbb{M}_\mathrm{L}=
-2\sum_{j=0}^{N}\ket{j}\bra{j}+
\sum_{j=0}^{N-1}(\ket{j+1}\bra{j}+\ket{j}\bra{j+1}),
\end{equation}
where, in contrast with the matrix in \eqref{eq03}, the matrix in \eqref{eq20} lacks of the ones at the corners, which maintain joined the array into a circle. The linear finite array implies that the chain of masses could have not material or physical constrictions on the borders; that is, just contour conditions to keep the system well defined. The set of coupled differential equations governing the evolution of this system is \cite{Meirovitch1986}
\begin{align}\label{eq21}
    \ddot{q}_0+k(2q_0-q_1)&=0,   \nonumber \\
    \ddot{q}_j+k(2q_j-q_{j+1}-q_{j-1})&=0, \\
    \ddot{q}_N+k(2q_N-q_{N-1})&=0. \nonumber
\end{align}
To solve this problem, we follow the presentation in \cite{rmf57.2} proposing the spectral decomposition of the interaction matrix as
\begin{equation}
\mathbb{M}_\mathrm{L}=\mathbb{S}\mathbb{D}\mathbb{S}^{-1},\label{eq22}
\end{equation}
where the matrix operator $\mathbb{S}$ is defined as
\begin{equation}
\mathbb{S}=\sum_{i,j=0}^{N}\frac{U_{i}(y_{j+1})}{\sqrt{\sum_{s=0}^{N}[U_{s}(y_{j+1})]^2}}\ket{i}\bra{j},\label{eq23}
\end{equation}
being $U_{i}(x)$ the Chebyshev polynomials of second kind \cite{Abramowitz1964,Olver2010}; $y_{j}=\cos(\phi_{j})$ with $\phi_{j}=\frac{j\pi}{N+2}, \; j=0,1,2,...,N$ as the $N+1$ roots of the polynomials. The matrix $\mathbb{D}$ is diagonal with their elements given by  \cite{Abramowitz1964,Olver2010}
\begin{equation}
\mathbb{D}=2k \sum_{j=0}^{N} \left[ \cos(\phi_{j+1})-1\right] \ket{j}\bra{j}.\label{eq24}
\end{equation}
Then, in analogy with Sec. \ref{circarray}, the solution of the matrix equation $\ddot{\mathbf{Q}}\left( t\right) =\mathbb{M}_\mathrm{L}$ is find in the factorization
\begin{equation}
\mathbf{Q}(t)=\mathbb{S}\cosh(t\sqrt{\mathbb{D}})\mathbb{S}^{-1}\mathbf{Q}(0).\label{eq25}
\end{equation}
Given the specific initial condition $\mathbf{Q}(0)=w\ket{l}$, with $l$ the number of the mass that is displaced with an amplitude $w$ respect to the equilibrium position, we arrive to the explicit form of the solution for every single mass $n$
\begin{equation}
q_n(t)=w\sum_{j=0}^{N}\frac{U_{n}(y_{j+1})U_{l}(y_{j+1})}{\sum_{s=0}^{N}[U_{s}(y_{j+1})]^2}\cos\left[2 t \sqrt{k}\sin\left( \frac{\phi_{j+1}}{2}\right) \right], \label{eq26}
\end{equation}
where it is worth to notice that the denominator is a sum of $N+1$ squared Chebyshev polynomials, that can be identified as a normalization dependence on the position of every single mass.
\begin{figure}[H]
    \begin{center}
        \includegraphics[width=\linewidth]{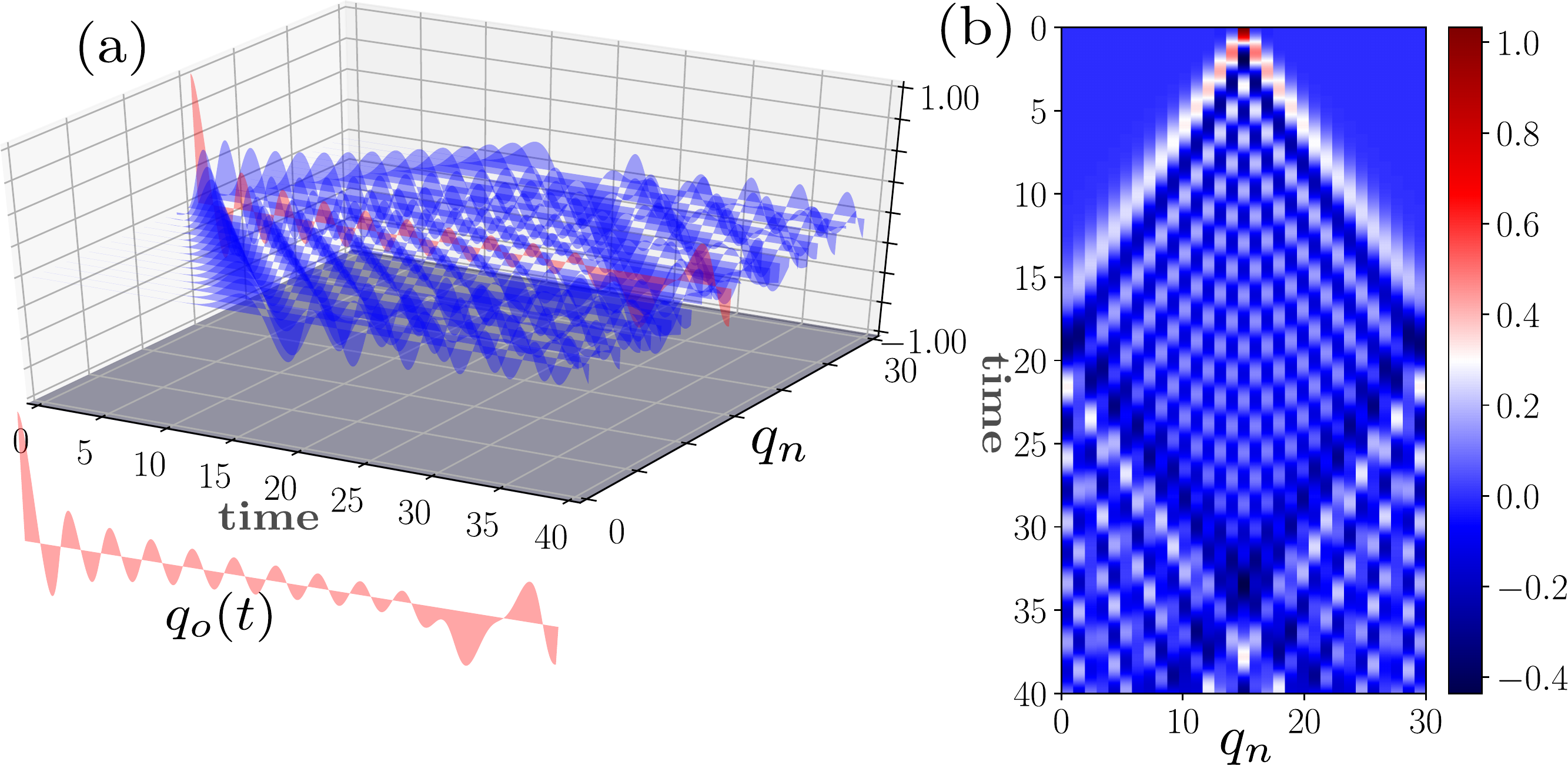}
    \end{center}
    \caption{Temporal evolution of the position of each mass in the linear finite array. The initial condition has unitary amplitude at $\mathbf{Q}(0)=\ket{15}$. We set $N=30$.}
    \label{fig4}
\end{figure}
In Figure \ref{fig4}, we plot the temporal evolution \eqref{eq26} for an initial condition $\mathbf{Q}(0)=\ket{15}$ with $N=30$; that is, $N+1$ elements in the array. A centralized initial condition is elected, because in the former system of equations \eqref{eq15} the boundary conditions forbids the interaction between the first and final elements of the array, so no amplitude is transmitted from $q_{0}$ to $q_{N}$. Setting the initial condition in the center of the array permits us observe that the propagation follows a symmetric evolution until it arrives to the edges, rebounds and recombine with the amplitudes coming from the center of the array, giving place to interference at times $t>20$. Of course, the points where the propagation meets the edges is a function of the pair values $\{m_{j},k_{j}\}$, because of our unit value election, the presented behavior follows.

\subsection{Kravchuk interaction}\label{linkr}
We can do a step forward and engineer an iteration matrix $\mat{M}_{L}$ which diagonalization is given in terms of hypergeometrical functions, more specific, discrete orthogonal polynomials. Following Regniers \cite{Regniers2009}, we can propose the interaction matrix to be of the form
\begin{equation}\label{eq27}
\mat{M}_{L} =
\begin{pmatrix}
- \alpha_{0} & \beta_{1}   & 0           & \cdots    & 0\\
  \beta_{1}  & -\alpha_{1} & \beta_{2}   & \cdots    & 0\\
  0          & \beta_{2}   & -\alpha_{2} & \ddots    & \vdots\\
  \vdots     & \vdots      & \ddots      & \ddots    & \beta_{N}\\
  0          & 0           & 0           & \beta_{N} & -\alpha_{N}\\
\end{pmatrix},
\end{equation}
where the coefficients follow the laws
\begin{align}\label{eq28}
\alpha_{i}=&Np+(1-2p)n,
\nonumber \\ 
\beta_{i}=&\sqrt{p(1-p)}\sqrt{i(N-i+1)}.
\end{align}
It can be proved that $p=1/2$ gives the diagonalization
\begin{equation}
\mat{M}_{L} = \mat{U}\mat{D}\mat{U}^{T},\label{eq29}
\end{equation}
where $\mat{D}=-\mathrm{diag}(0,1,\cdots,N)$ and the matrix elements of $\mat{U}$ are defined by
\begin{equation}
(\mat{U})_{i,j} = K_{i}(j) := \sqrt{\frac{w(j)}{h_{i}}}k_{i}(j),\label{eq30}
\end{equation}
with $w(j)=\binom{N}{j}p^{j}(1-p)^{N-j}$, and $h_{i}=\left(\frac{1-p}{p}\right)^{i}/\binom{N}{i}$. The functions $k_{i}(j)$ are the symmetric Kravchuk polynomials, whose use is extensive in the description of the discrete and finite harmonic oscillator \cite{ata,Weimann2016}. These polynomial are defined in terms of the Gaussian hypergeometric function as
\begin{equation}
k_{i}(j) := ~_{2}F_{1}(-j,-i,-N;2);\label{eq31}
\end{equation}
thus, the election of $p=1/2$ is justified for the obtainment of the symmetric functions that fulfill the requirements dictated by the spectral theorem \eqref{eq29}.\\
In order to solve the problem
\begin{equation}
\mathbf{Q}(t)=\cosh\left(t\sqrt{\mat{U}\mat{D}\mat{U}^{T}}\right)\mathbf{Q}(0),\label{eq32}
\end{equation}
we can write the matrices $\mat{U}$ and $\mat{D}$, in the discrete basis $\ket{l}$, as
\begin{equation}
\mat{U} = \sum\limits_{i,j=0}^{N} K_{i}(j) \ket{i}\bra{j},\qquad \mat{D} = -\sum\limits_{i=0}^{N} i\ket{i}\bra{i}.\label{eq33}
\end{equation}
As $\mat{D}$ is diagonal, we can obtain the proper factorization of $\mathbf{Q}(t)$ as in the previous cases in the form
\begin{equation}
\mathbf{Q}(t) =\mat{U}\cosh(t\sqrt{\mathbf{D}})\mat{U}^{T}\mathbf{Q}(0),\label{eq34}
\end{equation}
and then, with the definitions of the matrices \eqref{eq33} and the initial condition $\mathbf{Q}(0)=w\ket{l}$, we obtain the vector solution as
\begin{equation}
\mathbf{Q}(t)=w\sum\limits_{m,n=0}^{N}K_{m}(n)\cos(r\sqrt{n})K_{l}(n)\ket{m}.\label{eq35}
\end{equation}
Finally projecting over the $m$th element in the array, we obtain the single dynamical function for the position
\begin{equation}
q_{m} = w\sum\limits_{n=0}^{N}K_{m}(n)\cos(r\sqrt{n})K_{l}(n),\label{eq36}
\end{equation}
for $m\in\{0,1,\cdots,N\}$ and $l$ the index of the initial condition, that is, which mass is \emph{excited} at time $t=0$.\\
This last exercise was somehow easy because we can diagonalize $\mat{M}_{L}$ in terms of a pure diagonal matrix $\mat{D}$ and the \emph{Kravchuk matrices} $\mat{U}$. It is important to remark that the functions \eqref{eq30} are the solutions of the discretization of the quantum harmonic oscillator embedded in the compact algebra $\mathfrak{su}(2)$. This functions are a feasible approximation of the Hermite-Gauss functions in the discrete and finite space of the algebra.
\begin{figure}[H]
    \begin{center}
        \includegraphics[width=\linewidth]{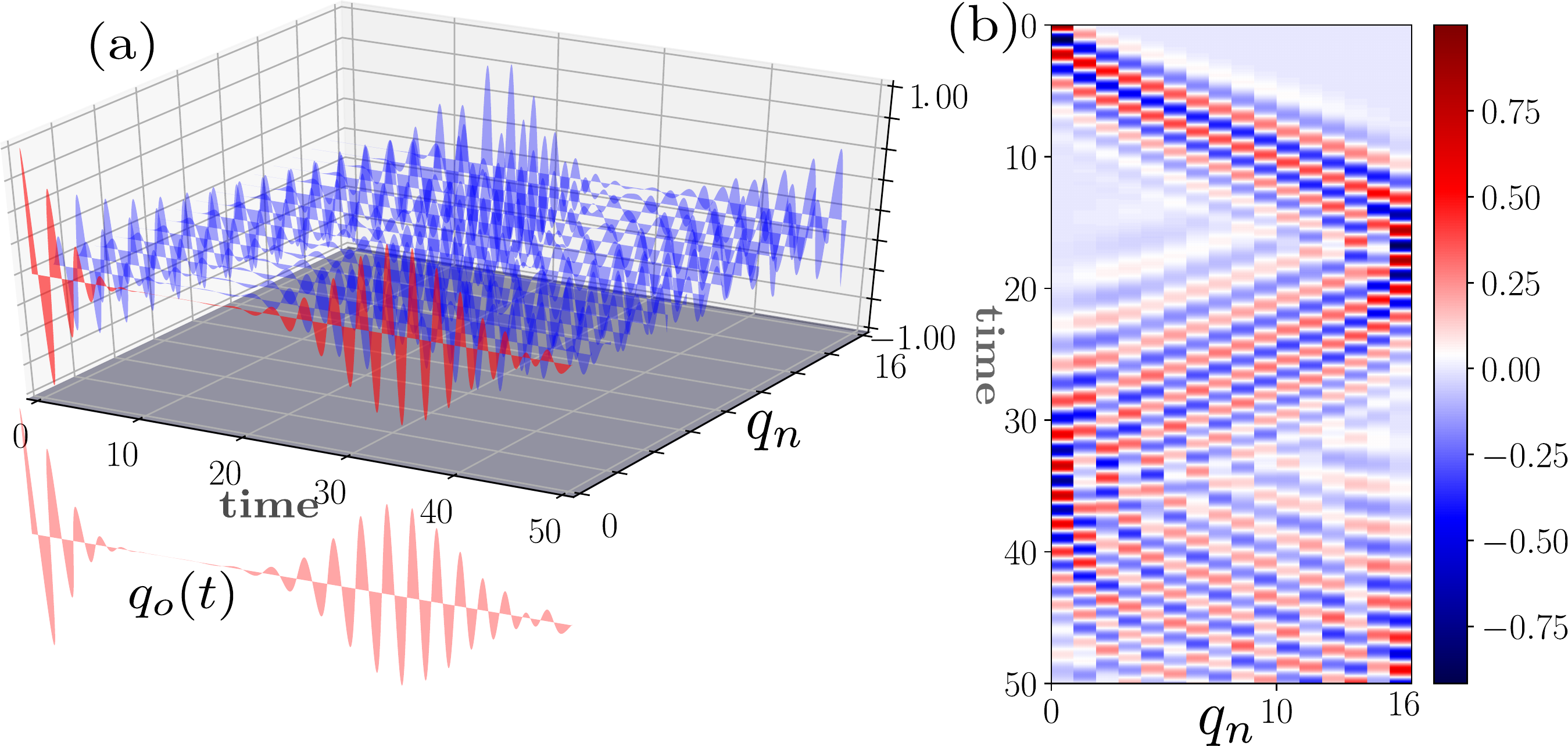}
    \end{center}
    \caption{Temporal evolution of the linear finite array when the interaction matrix is given by \eqref{eq27}. Here, we set $N=16$ and the initial condition with unit amplitude at $\mathbf{Q}(0)=\ket{0}$.}
    \label{fig5}
\end{figure}
In Figure \ref{fig5}, we plot the temporal evolution of \eqref{eq36}, here we set a dimension $N=16$ and an initial condition with unit amplitude at $\ket{0}$. We see that the slope of the propagation is nearly acute, giving that the initial excitation transmitted to the adjacent masses reach the edge at $q_{N}$ very rapidly, where a bouncing is observed; after that, the propagation follows another acute slope but with some recombination of amplitudes occurring. After the second bounce we observe not a regular pattern, just recombination and interference of amplitudes. This last behavior let us assume that the nearest neighbor interaction \eqref{eq27} presents some complex features after some time in the evolution \eqref{eq36}.

\subsection{Case $m_{j}=\binom{N}{N-j}$, $k_{j}=j m_{j}$}\label{linfac}
In this section, the masses $m_{j}$ and the restitution elements $k_{j}$ will be  defined in terms of the binomial coefficients. We propose to use
\begin{equation}
m_{j} = \binom{N}{N-j},\quad k_{j}=j m_{j},\qquad j\in\{0,1,\cdots,N\},\label{eq37}
\end{equation}
in such way that both quantities follow some sort of binomial distribution. It is important to notice that $m_{j}$ is well defined for all the values of $j$, but that the restitution elements has a null value when $j=0$. This last issue is not a problem nor a slip in the statements, it just says that the first restitution element is present with a intrinsic value of restitution equal to zero, following that the mathematical constraint is well defined. Physically this is the same picture of Figure \ref{figmk}, but the left and right material attachments are leave free.\\
Doing the calculations for the coefficients in \eqref{eq17}, we arrive to the definition of $\mathbb{M}_{L}$ as
\begin{align}\label{eq38}
\mathbb{M}_\mathrm{L} =& -N \sum_{j=0}^{N} \ket{j}\bra{j}+\sum_{j=0}^{N-1}\left(N-j \right)\ket{j}\bra{j+1} 
\nonumber \\ &
+ \sum_{j=0}^{N-1}\left( j+1 \right) \ket{j+1}\bra{j}.
\end{align}
The diagonal is constant with value $-N$ and the off-diagonals has up and downward values, from $1$ to $N$ and vice versa, respectively.\\
The coefficients previously showed can be cast to the operational representation as
\begin{align}\label{eq39}
\mathbb{K}_0&=\sum_{j=0}^{N}\left(\frac{N}{2}-j\right) \ket{j}\bra{j},\\
\mathbb{K}_+&=\sum_{j=0}^{N-1}\left(j+1\right) \ket{j}\bra{j+1},\\
\mathbb{K}_-&=\sum_{j=0}^{N-1}\left(N-j\right) \ket{j+1}\bra{j},
\end{align}
in such way that \eqref{eq38} can be rewritten as
\begin{equation}\label{eq40}
\mathbb{M}_\mathrm{L}=-N \mathbb{I}+\mathbb{K}_+ +\mathbb{K}_-.
\end{equation}
The operator given in \eqref{eq38} obeys the commutation relations
\begin{align}\label{eq41}
[\mathbb{K}_0,\mathbb{K}_-]=-\mathbb{K}_-,&\qquad
[\mathbb{K}_0,\mathbb{K}_+]=\mathbb{K}_+,
\nonumber \\
[\mathbb{K}_+,\mathbb{K}_-]&=2\mathbb{K}_0;
\end{align}
thus, they are a representation of the Lie algebra $\mathfrak{su}(2)$. Using the commutation relations, we transforms the operators \eqref{eq38} in such way that
$\mathbb{M}_{L}$ is diagonal. It is long, but straightforward, to arrive to the set of transformations
\begin{subequations}
    \begin{align}\label{eq42}
    &\exp\left( \alpha \mathbb{K}_- \right)
    \mathbb{K}_+
    \exp\left( -\alpha \mathbb{K}_- \right)
    =\mathbb{K}_+ -2 \alpha \mathbb{K}_0-\alpha^2 \mathbb{K}_-,
    \\
    &\exp\left( \alpha \mathbb{K}_+ \right)
    \mathbb{K}_-
    \exp\left( -\alpha \mathbb{K}_+ \right)
    =\mathbb{K}_- + 2\alpha \mathbb{K}_0-\alpha^2 \mathbb{K}_+,
    \\
    &\exp\left( \alpha \mathbb{K}_+ \right)
    \mathbb{K}_0
    \exp\left( -\alpha \mathbb{K}_+ \right)
    =\mathbb{K}_0 - \alpha \mathbb{K}_+,
    \\
    &\exp\left( \alpha \mathbb{K}_- \right)
    \mathbb{K}_0
    \exp\left( -\alpha \mathbb{K}_- \right)
    =\mathbb{K}_0 + \alpha \mathbb{K}_-.
    \end{align}
\end{subequations}
Hence, the interactions matrix can be written as
\begin{align}\label{eq43}
e^{\beta \mathbb{K}_+}&e^{\alpha \mathbb{K}_-}\mathbb{M}_\mathrm{L}e^{-\alpha \mathbb{K}_-}e^{-\beta \mathbb{K}_+}=
 \nonumber \\ &
=-N \mathbb{I}+2\left(-\alpha^2 \beta-\alpha+\beta \right) \mathbb{K}_0
\nonumber \\ &
+\left(\alpha^2\beta^2+2\alpha\beta-\beta^2+1 \right) \mathbb{K}_++\left(1-\alpha^2 \right)\mathbb{K}_-.
\end{align}
This last equation can be carry to a full diagonal form if we choose the parameters to be $\alpha=-1$ and $\beta=1/2$,
\begin{equation}\label{eq44}
e^{\beta \mathbb{K}_+}e^{\alpha \mathbb{K}_-}\mathbb{M}_\mathrm{L}e^{-\alpha \mathbb{K}_-}e^{-\beta \mathbb{K}_+} =
-2 \Lambda,
\end{equation}
where $\Lambda=\sum_{j=0}^{N}j\ket{j}\bra{j}$. Inverting \eqref{eq44}, we finally obtain that
\begin{equation}\label{eq45}
\mathbb{M}_\mathrm{L}=-2e^{\mathbb{K}_-}e^{ -\mathbb{K}_+/2}\Lambda e^{ \mathbb{K}_+/2}e^{-\mathbb{K}_-}.
\end{equation}
The diagonal representation of $\mathbb{M}_{L}$ given above, let us now look for the solution of the initial value problem \eqref{eq19} as
\begin{equation}\label{eq46}
\mathbf{Q}(t)=e^{\mathbb{K}_-}e^{ -\mathbb{K}_+/2}\cos\left( t \sqrt{2\Lambda}\right)  e^{ \mathbb{K}_+/2}e^{-\mathbb{K}_-}\mathbf{Q}(0).
\end{equation}
\begin{figure}[H]
    \begin{center}
        \includegraphics[width=\linewidth]{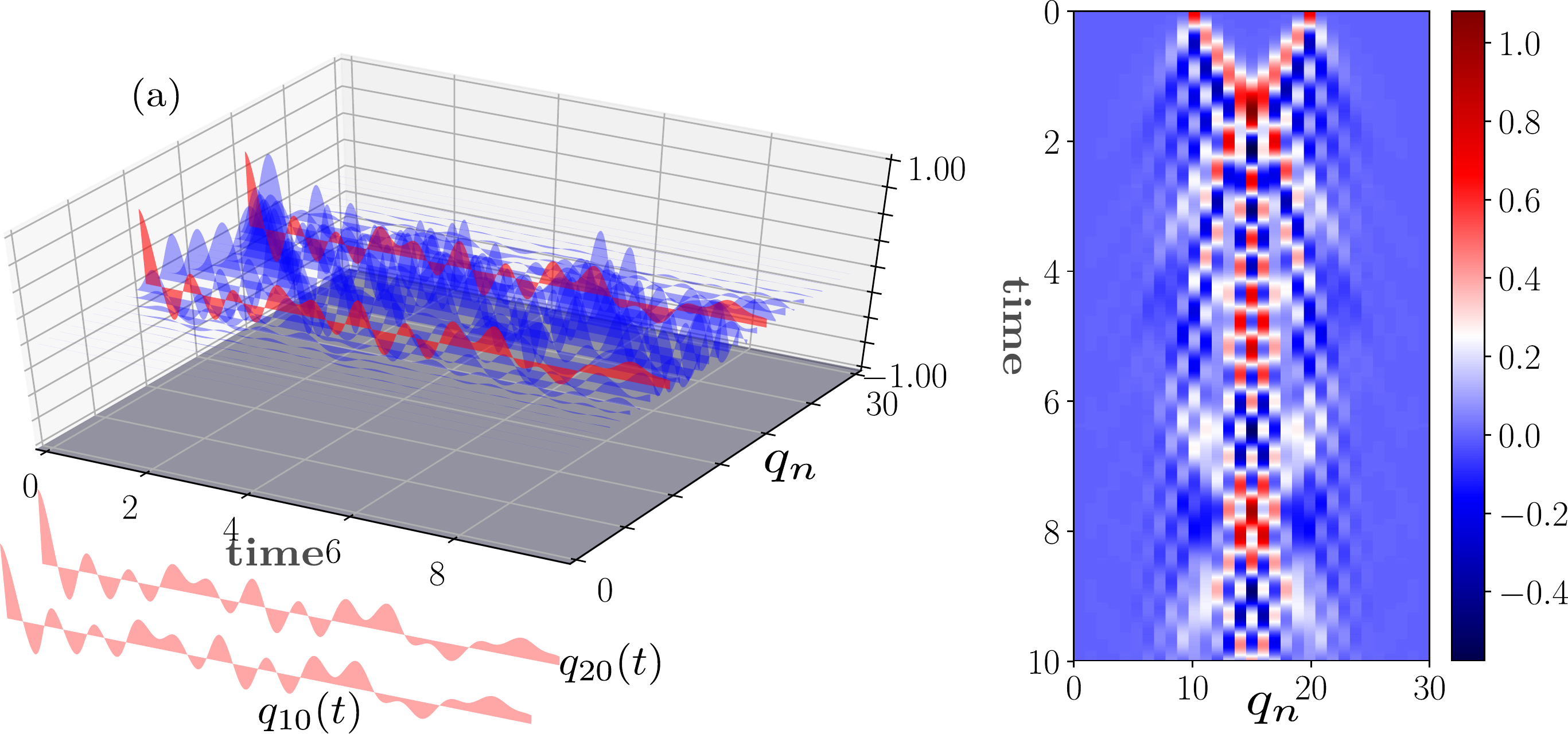}
    \end{center}
    \caption{Temporal evolution of the linear finite array when the interaction matrix is modeled by \eqref{eq38}. We set the dimension $N=30$ and the initial condition as the coherent superposition $\mathbf{Q}(0)=1/2(\ket{10}+\ket{20})$.}
    \label{fig6}
\end{figure}
Expression \eqref{eq46} is an analytic closed expression; however, their explicit calculation is cumbersome. Therefore, instead of giving the long-complete expression, we accelerate the process using the matrix representation \eqref{eq39} for a fixed $N$ and make a numeric evaluation for the initial condition $\mathbf{Q}(0)=w\ket{l}$. Also due to the definition of the masses and restitution elements, when we excite masses near $j=0$ the dynamics is not interesting, because around this specific place the restitution constant vanish. So, to obtain relevant results in the dynamics, we propose to use initial conditions around the middle mass. Figure \ref{fig6} shows a numerical evaluation of \eqref{eq46} when the initial condition is a coherent state around the center of the array.

\section{Conclusions}
In this work we present a set of mechanical systems that are
solved using mathematical methods and arguments often encountered
in the analysis and solution of quantum optical phenomena. We
found that the time evolution of the position amplitude of a
chain of masses has some resemblance with light propagation in graded indexed waveguide arrays; we may conclude that in some particular cases (specific values of spring constants and masses) there is an isomorphism between both systems.  This conclusion may be relevant because it motivates the search of relations between mechanical systems of coupled oscillators and systems that obey Schr\"odinger and Helmholtz-like equations \cite{Unpublished}. Finally, we may say that the use of these quantum optical methods are fully equivalent to those that follow Hamiltonian or Lagrangian developments for many-body interactions. The simplicity of the solution arise from the fact that the interaction matrix is fully diagonalizable.

\section{Acknowledgments}
Alejandro R.U. acknowledge CONACyT for their financial support in the development of this work through Ph.D. grant \#449192. I.R.-P. thanks Prof. J. R\'ecamier for his hospitality at ICF-UNAM and acknowledge partial support from DGAPA UNAM project PAPIIT IN111119, and Beca de Colaboraci\'on INAOE.


\begin{thebibliography}{10}

    \bibitem{Meirovitch1986}
    L.~Meirovitch, {\em Elements of vibration analysis}.
    \newblock McGraw-Hill, second edition~ed., 1986.

    \bibitem{landauM}
    E.~L. L.D.~Landau, {\em Course of Theoretical Physics: Vol. 1, Mechanics}.
    \newblock Butterworth-Heinemann, 1976.

    \bibitem{Kotkin1980}
    G.~Kotkin and V.~Serbo, {\em Problemas de mec\'anica cl\'asica}.
    \newblock Editorial MIR, 1980.

    \bibitem{YaFrenkel1938}
    Y.~I. Frenkel and T.~Kontorova {\em Phys. Z. Sowietunion}, vol.~13, p.~1, 1938.

    \bibitem{T.A.Kontorova1938}
    T.~A. Kontorova and Y.~I. Frenkel {\em Zh. Eksp. Teor. Fiz.}, vol.~8, p.~89,
    1938.

    \bibitem{T.A.Kontorova1938a}
    T.~A. Kontorova and Y.~I. Frenkel {\em Zh. Eksp. Teor. Fiz.}, vol.~8, p.~1340,
    1938.

    \bibitem{Braun1998}
    O.~M. Braun and Y.~S. Kivshar, ``Nonlinear dynamics of the Frenkel-Kontorova
    model,'' {\em Physics Reports}, vol.~306, pp.~1--108, 1998.

    \bibitem{Allen1998}
    P.~B. Allen and J.~Kelner, ``Evolution of a vibrational wave packet on a
    disordered chain,'' {\em Am. J. Phys.}, vol.~66, p.~497, June 1998.

    \bibitem{Makris2006}
    K.G. Makris, and D.~N.
    Christodoulides, ``Method of images in optical discrete systems,'' {\em Physical Review E}, vol.~73,
    pp.~036616, 2006.


    \bibitem{Keil2011}
    R.~Keil, A.~Perez-Leija, F.~Dreisow, M.~Heinrich, H.~Moya-Cessa, S.~Nolte,
    D.~N. Christodoulides, and A.~Szameit, ``Classical analogue of displaced fock
    states and quantum correlations in glauber-fock photonic lattices,'' {\em
        Physical Review Letters}, vol.~107, aug 2011.


    \bibitem{PerezLeija2010}
    A.~Perez-Leija, R. Keil,  A.~Szameit, A. Abouraddy, H.~Moya-Cessa, and D.~N. Christodoulides,
    ``Tailoring the correlation and anti-correlation behavior of path-entangled photons in Glauber-Fock  oscillator lattices,'' {\em Physical Review A},
    vol.~55, p.~013848, 2012.

    \bibitem{PerezLeija2016}
    A.~Perez-Leija, L.~A. Andrade-Morales, F.~Soto-Eguibar, A.~Szameit, and H.~M.
    Moya-Cessa, ``The Pegg{\textendash}Barnett phase operator and the discrete
    Fourier transform,'' {\em Physica Scripta}, vol.~91, p.~043008, mar 2016.

    \bibitem{PrezLeija2012}
    A.~P{\'{e}}rez-Leija, H.~Moya-Cessa, and D.~N. Christodoulides, ``Optical
    realization of the atom{\textendash}field interaction in waveguide
    lattices,'' {\em Physica Scripta}, vol.~T147, p.~014023, feb 2012.

    \bibitem{Atakishiyev2008}
    N.~M. Atakishiyev, A.~U. Klimyk, and K.~B. Wolf, ``A discrete quantum model of
    the harmonic oscillator,'' {\em Journal of Physics A: Mathematical and
        Theoretical}, vol.~41, p.~085201, feb 2008.

    \bibitem{ata}
    N.~Atakishiyev, S.~Pogosyan, and K.~Wolf, ``Finite models of the oscillator,''
    {\em Physics of Particles and Nuclei}, vol.~36, pp.~247--265, 05 2005.

    \bibitem{London1926}
    F.~London {\em Z. Phys.}, vol.~37, pp.~915--925, 1926.

    \bibitem{moya2011differential}
    H.~M. Moya-Cessa and F.~Soto-Eguibar, {\em Differential equations: an
        operational approach}.
    \newblock Rinton Press, 2011.

    \bibitem{MoyaCessa2018}
    H.~M. Moya-Cessa and F.~Soto-Eguibar, ``Discrete fractional Fourier transform:
    Vandermonde approach,'' {\em {IMA} Journal of Applied Mathematics}, jun 2018.

    \bibitem{rmf57.2}
    F.~Soto-Eguibar, O.~Aguilar-Loreto, A.~Perez-Leija, H.~Moya-Cessa, and
    D.~Christodoulides, ``Finite photonic lattices: a solution using
    characteristic polynomials,'' {\em Revista Mexicana de F\'isica}, vol.~57,
    pp.~158--161, April 2011.

    \bibitem{Abramowitz1964}
    M.~Abramowitz and I.~A. Stegun, {\em Handbook of Mathematical Functions with
        Formulas, Graphs, and Mathematical Tables}.
    \newblock Dover, 1964.

    \bibitem{Olver2010}
    F.~W.~J. Olver, , D.~W. Lozier, R.~F. Boisvert, and C.~W. Clark, {\em The
        {NIST} Handbook of Mathematical Functions}.
    \newblock Cambridge Univ. Press, 2010.

    \bibitem{Regniers2009}
    G.~Regniers and J.~V. der Jeugt, ``Analytically solvable hamiltonians for
    quantum systems with a nearest-neighbour interaction,'' {\em Journal of
        Physics A: Mathematical and Theoretical}, vol.~42, p.~125301, feb 2009.

    \bibitem{Weimann2016}
    S.~Weimann, A.~Perez-Leija, M.~Lebugle, R.~Keil, M.~Tichy, M.~Gr{\"a}fe,
    R.~Heilmann, S.~Nolte, H.~Moya-Cessa, G.~Weihs, D.~N. Christodoulides, and
    A.~Szameit, ``Implementation of quantum and classical discrete fractional
    fourier transforms,'' {\em Nature Communications}, vol.~7, pp.~11027 EP --,
    Mar 2016.

    \bibitem{Unpublished} A. R. Urz\'ua, I. Ramos-Prieto, F.
    Soto-Eguibar, V. Arriz\'on, and H.M. Moya-Cessa, unpublished.
    \newblock Article.
\end{thebibliography}
\end{document}